\documentclass[twocolumn,amsmath,amssymb,prl]{revtex4}
\usepackage{latexsym}
\usepackage{graphicx}
\usepackage{psfig}
\usepackage{color}
\usepackage[colorinlistoftodos]{todonotes}
\usepackage{verbatim}

\usepackage[letterpaper, margin=1in]{geometry}

\begin{document}

\title{Tuning supercurrent in Josephson field effect transistors using h-BN dielectric}

\author{Fatemeh Barati$^{1}$}
\altaffiliation{These authors contributed equally.}

\author{Josh P.~Thompson$^{2}$}
\altaffiliation{These authors contributed equally.}

\author{Matthieu C. Dartiailh$^{1}$}
\altaffiliation{These authors contributed equally.}

\author{Kasra Sardashti$^{1}$}

\author{William Mayer$^{1}$}

\author{Joseph Yuan$^{1}$}

\author{Kaushini Wickramasinghe$^{1}$}

\author{Kenji Watanabe$^{3}$}

\author{Takashi Taniguchi$^{4}$}

\author{Hugh Churchill$^{2}$}

\author{Javad Shabani$^{1}$}

\affiliation{$^{1}$Center for Quantum Phenomena, Department of Physics, New York University, NY 10003, USA
\\
$^{2}$ Department of Physics, University of Arkansas, Fayetteville, Arkansas 72701, USA
\\
$^{3}$ Research Center for Functional Materials, National Institute for Materials Science, 1-1 Namiki, Tsukuba 305-0044, Japan
\\
$^{4}$ International Center for Materials Nanoarchitectonics, National Institute for Materials Science, 1-1 Namiki, Tsukuba 305-0044, Japan
}

\begin{abstract}
Epitaxial Al-InAs heterostructures appear as a promising materials platform for exploring mesoscopic and topological superconductivity. A unique property of Josephson Junction Field Effect Transistors (JJ-FETs) fabricated on these heterostructures is the ability to tune the supercurrent using a metallic gate. Here we report the fabrication and measurement of gate-tunable Al-InAs JJ-FETs in which the gate dielectric in contact with the InAs is produced by mechanically exfoliated hexagonal boron nitride (h-BN) followed by dry transfer. We discuss a versatile fabrication process that enables compatibility between layered material transfer and Al-InAs heterostructures that allows us to achieve full gate-tunablity of supercurrent by using only 5~nm thick h-BN flakes. Our study shows that pristine properties of epitaxial Josephson junctions, such as product of normal resistance and critical current, I$_{\rm c}$R$_{\rm n}$, are preserved. Furthermore complementary measurements confirm that using h-BN dielectric changes the channel density less when compared to atomic layer deposition of Al$_2$O$_3$.
\end{abstract}

\maketitle

Understanding and engineering JJ-FETs fabricated on semiconductors with highly transparent contacts can yield a gate-controllable supercurrent \cite{mayer,Lee2019, mayer2019anom,clark_feasibility_1980,richter_transport_1999,kroemer_electronic_1994,heida_nonlocal_1998,akazaki_josephson_1996,calado_ballistic_2015}. Tuning the conductivity of the semiconductor part of a JJ-FET affects superconducting properties.  For example, JJ-FETs fabricated on Al-InAs have been used for tunable superconducting qubits, the so-called ``gatemon'', where the qubit frequency, which depends on the value of the supercurrent, can be tuned in-situ with an applied electric field \cite{Casparis2018}. Furthermore, since InAs has large spin-orbit coupling, Al-InAs system can host topological superconductivity and Majorana bound states \cite{mayer2019phase,Fornieri2019,suominen_zero-energy_2017,nichele_scaling_2017}.  In this work we discuss a novel fabrication process for gating semiconductor-based JJ-FETs with 2D h-BN dielectric to study their quantum transport properties. This hybrid approach can reduce the unintentional doping of JJ-FETs similar to earlier work in 2D materials using h-BN encapsulation in graphene \cite{dean_boron_2010} and MoS2 devices \cite{Lee2015}. 

\begin{figure*}
\begin{center}
\includegraphics[width=1\textwidth]{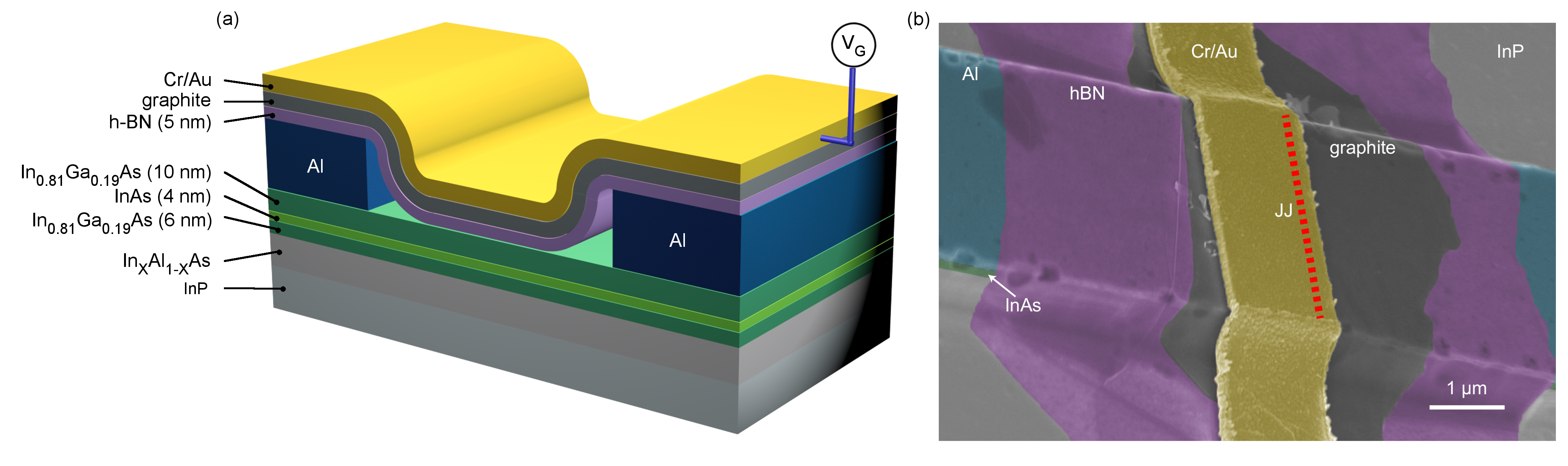}
\caption{(a) Schematic of the gated JJ-FET geometry. The gate stack comprises a thin layer of h-BN as gate dielectric with Cr/Au-contacted graphite as the gate electrode. (b) Scanning electron microscopy image of the JJ-FET. The location of the JJ-FET channel is shown by the red dashed line. }
\label{fig:Figure1}
\end{center}
\end{figure*}

JJ-FET devices on III-V materials above the superconducting critical temperature behave practically like a III-V transistor. Historically, performance enhancement of these devices has been achieved through shrinking the gate length of transistors. However, the constant electrostatic scaling rule was not sufficient when the gate lengths of transistors reached the nanoscale. To revive the voltage scaling in order to reduce the power consumption of the integrated circuits new materials for the gate dielectric were explored. The advent of high-k dielectrics offered one encouraging solution. However it was found out that special care had to be taken to produce devices with sufficiently low density of charge traps at the interface of surface channel and dielectric for reliable device operation. While the standard atomic layer deposition of high-k oxides e.g. Al2O3 is widely used and optimized, the research for minimizing the trap states on III/V materials continues \cite{chobpattana_nitrogen-passivated_2013, shahrjerdi_impact_2007,shahrjerdi_hall_2010, burek_influence_2011}.

The ability to pick and place an ultra-thin insulator flake with no chemical bonding to the InGaAs/InAs channel can minimize the trap states at the interface. In addition small flakes could be very useful when Al$_2$O$_3$ lift off is impractical. This can be used for example as a solution for using tunable elements in microwave circuits  and Gatemon qubits \cite{Casparis2018, casparis_voltage-controlled_2019, wang_quantum_2018, sardashti_voltage-tunable_2020}. Among two-dimensional (2D) materials, h-BN, with a bandgap of 5.9 eV, has been widely used as an insulating layer. It has dangling-bond-free surface, low density of charge traps, high surface optical phonon frequency, and low microwave absorption \cite{jang,huang_versatile_2020,alharbi_effect_2019,Wang2019,jekat_exfoliated_2020}. In this letter we integrate h-BN as a gate dielectric for epitaxial Al/InAs JJ-FETs. 

The InAs heterostructures are grown on semi-insulating InP (100) substrates using a modified Gen II molecular beam epitaxy machine. The step graded buffer layer, In$_{x}$Al$_{1-x}$As, is grown at low temperature to minimize dislocations formation due to the lattice mismatch between the active region and the InP substrate \cite{shabani_gating_2014, shabani_apparent_2014}. The quantum well consists of a 4 nm layer of InAs grown on a 6 nm layer In$_{0.81}$Ga$_{0.19}$As. The surface InAs quantum wells have relatively low electron mobilities mostly due to the direct contact of electrons to scattering impurities at the surface. A top InGaAs barrier can improve the situation by separating the quantum well from the Al interface. InGaAs is a suitable choice because it provides a soft confinement allowing sufficient overlap of electrons with superconductor for proximity effect as well as moving the wave-function away form the surface resulting in a mobility increase \cite{wickramasinghe_transport_2018}. In this study, 10 nm of In$_{0.81}$Ga$_{0.19}$As, is grown on InAs strained quantum well. After the quantum well is grown, the substrate is cooled to below zero centigrade to promote the growth of epitaxial Al (111) \cite{shabani_two-dimensional_2016}.

The JJ-FET fabrication process is performed by electron beam lithography using polymethyl methacrylate (PMMA) as the resist. Device mesas are defined and etched using Transene type D Al etchant followed by a III-V etch solution (see Supporting information (SI) for additional details). In the active region of the junction, the mesa is 4 $\mu$m wide. In a second step, the gap between superconducting contacts of the JJ was patterned by selective etching of Al over InGaAs/InAs using Transene type D. All presented samples have a gap of $\sim$100 nm.  To form the gate stack, h-BN and graphite flakes were mechanically exfoliated from bulk crystals onto 90 nm SiO$_2$/Si substrates at 100 $^{\circ}$C using a heated exfoliation method. Using an optical microscope, we selected appropriate graphite ($\sim$30 nm) and thin h-BN ($\sim$6 nm) flakes based on optical contrast and thickness uniformity. The gate stack was then transferred onto the JJ and finally contacted on top by Cr/Au.  Figure~1(a) illustrates the JJ-FET schematics from substrate to the gate stack using h-BN. Voltage (V$_{\rm G}$) can be applied to the graphite top gate (gray in Fig.~1(a)) which controls the carrier density of the InAs surface channel. Figure~1(b) shows a scanning electron microscope image of the JJ-FET device with false color representing the schematic layers. 

During initial device fabrication attempts, a stamp made of a thin polycarbonate (PC) film on polydimethylsiloxane (PDMS) was used to construct the graphite/h-BN stack and transfer it onto aluminum contacts.  However, the chloroform used to remove the PC after transfer etched the aluminum contacts and destroyed the Al/InAs junctions, likely because of a small concentration of HCl created in the chloroform by a reaction with oxygen. To eliminate chloroform from the fabrication process, a modified version of this technique was used to create stamps with polypropylene carbonate (PPC), which was dissolved in anisole. 
\begin{figure}[ht!]
\includegraphics[width=0.47\textwidth]{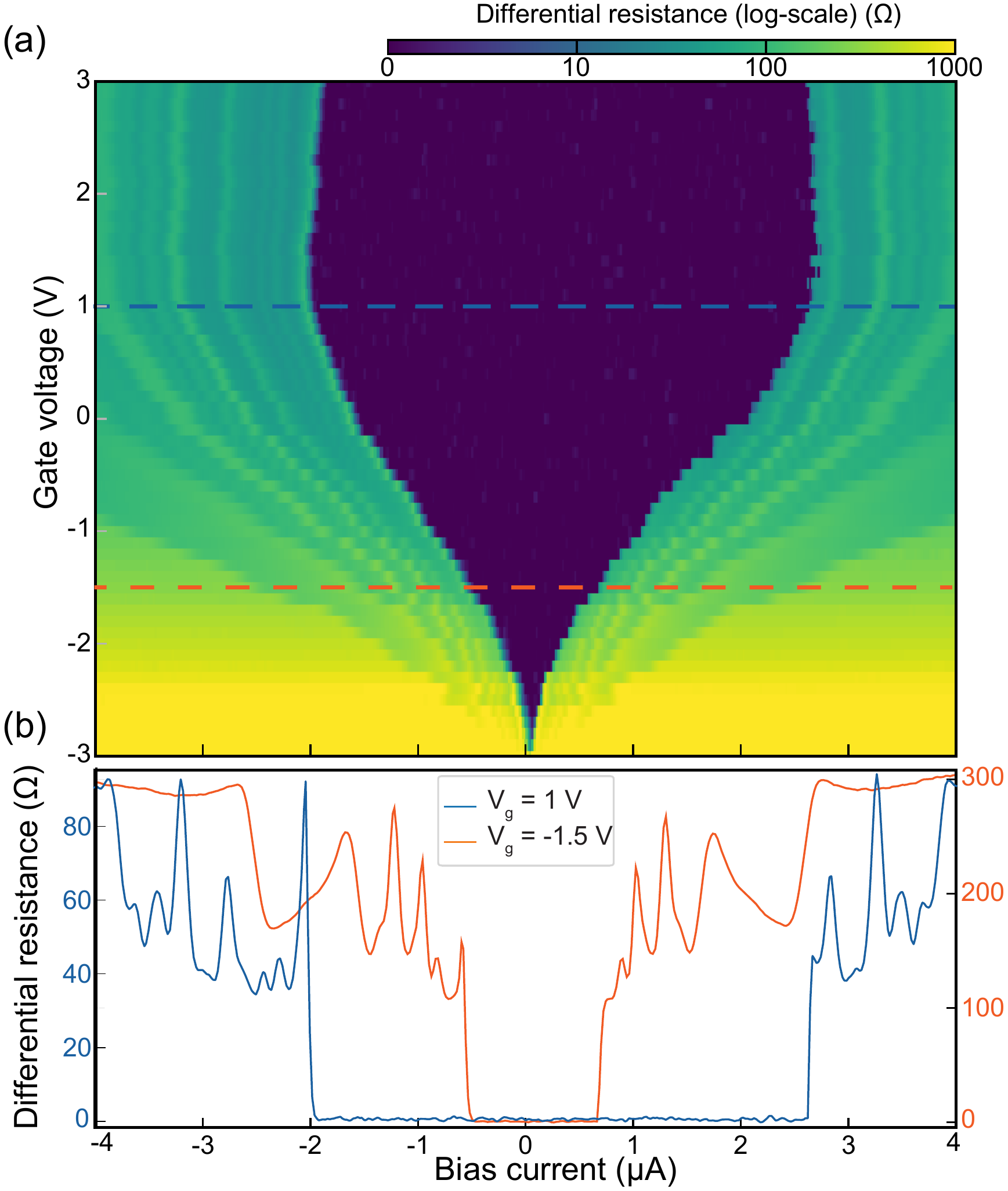}
\caption{(a) Differential resistance (dV/dI) of the JJ as a function of bias current and gate voltage. (b) Two line traces of panel (a) at V$_{\rm G}$ = 1 and -1.5V shown in blue and red, respectively.}
\label{fig:Figure2}
\end{figure}

Here, we focus on two JJ-FET devices A and B fabricated on nominally identical heterostructures. The only difference between these two samples is the gate dielectric. In sample A, 6 nm of hBN are used as the gate dielectric while sample B has 50 nm of Al$_2$O$_3$ instead. The 2DEGs in these heterostructures were characterized separately by magneto-resistance measurements in van der Pauw geometry. Both samples have density of 7 $\times$ 10$^{11}$ cm$^{-2}$. They also have an estimated mean free path $l_e$ of 200 nm. As a consequence with superconducting electrode gap of $L = $100~nm both samples are expected to be close to the ballistic regime $L < l_e$. The superconducting gap of the Al was estimated to be about 210 $\mu$eV from the critical temperature of the film in both samples ($T_c \sim 1.4$~K). In both samples, the Thouless energy $E_{\textrm{Thouless}} = \frac{\hbar\,v_F\,l_e}{2\,L^2}$, with $v_F$ the Fermi velocity and L the gap of the JJ, is larger than 1 meV, which implies that our junctions are in the short limit ($\Delta \ll E_{\textrm{Thouless}}$).

\begin{figure*}[ht!]
\includegraphics[width=0.98\textwidth]{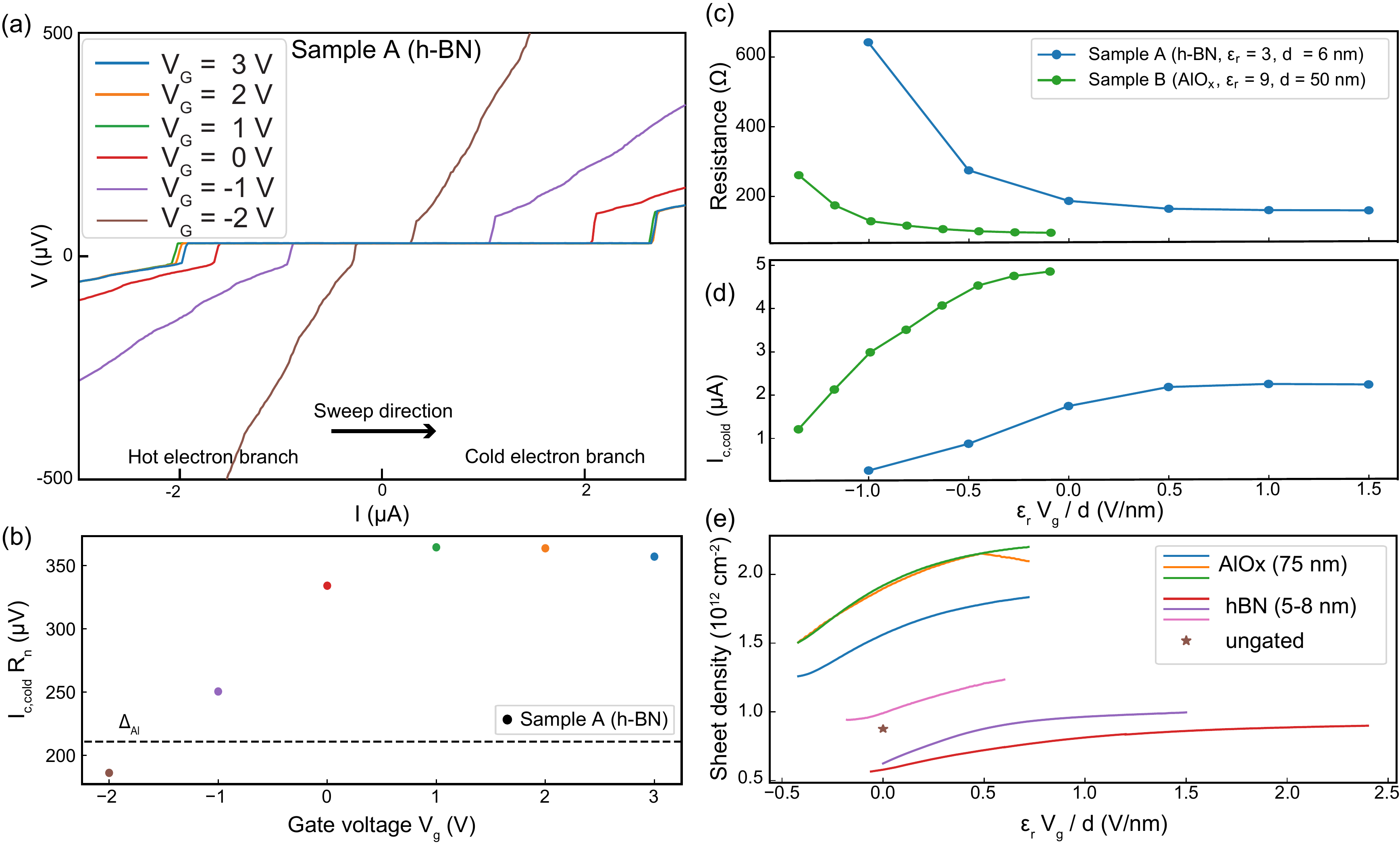}
\caption{(a) Voltage-current curve as a function of gate voltages from 3 V to -2 V in 1 V steps. (b) I$_{\rm c, cold}$R$_{\rm n}$ as a function of gate voltage for sample A (c) R$_{\rm n}$ as function of gate voltage for sample A with h-BN and sample B with Al$_2$O$_3$ dielectrics. (d) I$_{\rm c, cold}$ as function of gate voltage for samples A with h-BN, and sample B with an Al$_2$O$_3$ dielectrics.}
\label{fig:Figure3}
\end{figure*}

Figure~2a shows the differential resistance (dV/dI) vs.~measured d.c.~bias current as a function of applied top gate voltage at 30 mK.  We observed that the critical current remains nearly unchanged above V$_{\rm G}$ $>$ 1 V while it decreases below this value similar to  JJ-FET devices with Al$_2$O$_3$ as the gate dielectric \cite{mayer}. As we tune the gate voltage toward more negative values, the window with zero resistance state (dark blue) becomes narrower and reaches zero at V$_{\rm G}$= -3 V. Figure~2b shows the differential resistance as a function of bias current extracted from panel (a). The line traces in blue and red are for V$_{\rm G}$ = 1 V and -1.5 V, respectively. The critical current shows an asymmetric behavior that can be associated to the JJ hysteresis (see SI). The critical current is consistently higher on the cold branch where the cold branch goes from zero bias to high bias and corresponds to a lower effective electronic temperature before the transition out of the superconducting state. For example, at V$_{\rm G}$ = 1 V exhibiting -2 $\mu$A on the hot branch and 2.7 $\mu$A on the cold branch. We observe several peaks and valleys in the region above the critical current. In SI, we plot the conductance as a function of the voltage which allows us to interpret those as multiple Andreev reflection processes up to 4th order \cite{Morten2017, pankratova_multi-terminal_2018}. These observations support a highly transparent boundary with the superconducting contacts. In each Andreev reflection, two electrons are transferred across the junction and contribute to the current to drop the resistance, with peaks in resistance corresponding to transitions between successive orders of reflection processes in the high-transparency limit\cite{Morten2017}. These reflections are observed across the whole range of gate voltages and only disappear around V$_{\rm G}$ = -2.5 V, just above when supercurrent in the JJ ceases, V$_{\rm G}$ = -3 V.

Figure~3a shows the d.c.~transport measurement of voltage versus bias current as a function of top gate voltage. As the gate is used to deplete the 2DEG, by tuning the gate voltage from 3 V toward -2 V (color coded from blue to brown), the voltage increases and the critical current decreases. A figure of merit of JJ devices is considered to be their product of critical current multiplied by their normal state resistance. Theoretically this product is related to the induced superconducting gap and can reach $\pi$ for a ballistic junction with perfect interfaces \cite{ambegaokar_tunneling_1963,likharev_superconducting_1979}.  Figure~3b illustrates the gate dependence of the I$_{\rm c}\,$R$_{\rm n}$ products for sample A. The negative voltage trend is very similar to JJ devices with Al$_2$O$_3$ dielectrics \cite{mayer}. Data presented in SI suggests that the induced gap does not vary widely with gate and is close to the parent Al gap, which allows us to evaluate the maximum of the product I$_{\rm c}\,$R$_{\rm n}/\Delta_{\text{Al}} \sim$ 1.7. While we cannot resolve changes in the induced gap, high order MAR appear to be affected by gate voltage suggesting that the decrease of the I$_{\rm c}\,$R$_{\rm n}$ product we observe may be imparted both to the reduced mobility of the 2DEG at low density and to reduced transparency between the gated and Al-covered sections of the 2DEG.

For comparison we provide data from sample B with 50~nm of Al$_2$O$_3$ dielectric. Figure~3c makes a comparison between gate dependence of normal resistance (at high current bias) of two JJ-FETs with h-BN (sample A) and Al$_2$O$_3$ (sample B) dielectric. Sample B shows a smaller normal resistance compared to Sample A which suggests a higher density for Sample B. The gate efficiency of the device defined by $\delta Ic/\delta V_{G}$, can also be evaluated by plotting supercurrent as a function of electric displacement field (to normalize for different thicknesses and dielectric constants of h-BN and Al$_2$O$_3$) which is shown in Fig.~3d. Although the supercurrent is smaller in the h-BN device by a factor of two, the overall I$_{\rm c}\,$R$_{\rm n}$ are closer (within 30\%) due to higher R$_{\rm n}$. In terms of efficiency while h-BN can be made a few monolayers, the low dielectric constant of h-BN make the gate efficiency similar to Al$_2$O$_3$. While the curves overall look similar the shift in electric field is evident. At zero gate bias, the device with Al$_2$O$_3$ dielectric places the curve in saturation (as seen in Fig. 3c and d) while the device with h-BN dielectric is away from this region. This could be the results of higher density in devices with Al$_2$O$_3$ dielectric. To test whether Al$_2$O$_3$ deposition and h-BN modify the carrier density of 2DEGs we have made multiple small Hall bars and have deposited Al$_2$O$_3$ and transferred larger flakes of h-BN on them. We have described details of the Hall bar samples in Supporting Information. The summary of our study is shown in Fig.~3e. For reference, we have also added a data point from a Hall bar with no dielectric noted as ungated, which match another measurement made on an ungated Van der Pauw. While there are some variations, devices with Al$_2$O$_3$ dielectric show an increase in density compared to ungated reference sample up to two-fold. On the other hand, the devices with h-BN dielectric show smaller density change some above and some below the reference sample. This confirms our hypothesis that h-BN introduces less doping through interface states compared to Al$_2$O$_3$. This becomes an important factor in study of device physics and specially in study of topological superconductivity where interfaces play a crucial role.

\begin{figure}
\label{figure4}
\includegraphics[width=0.47\textwidth]{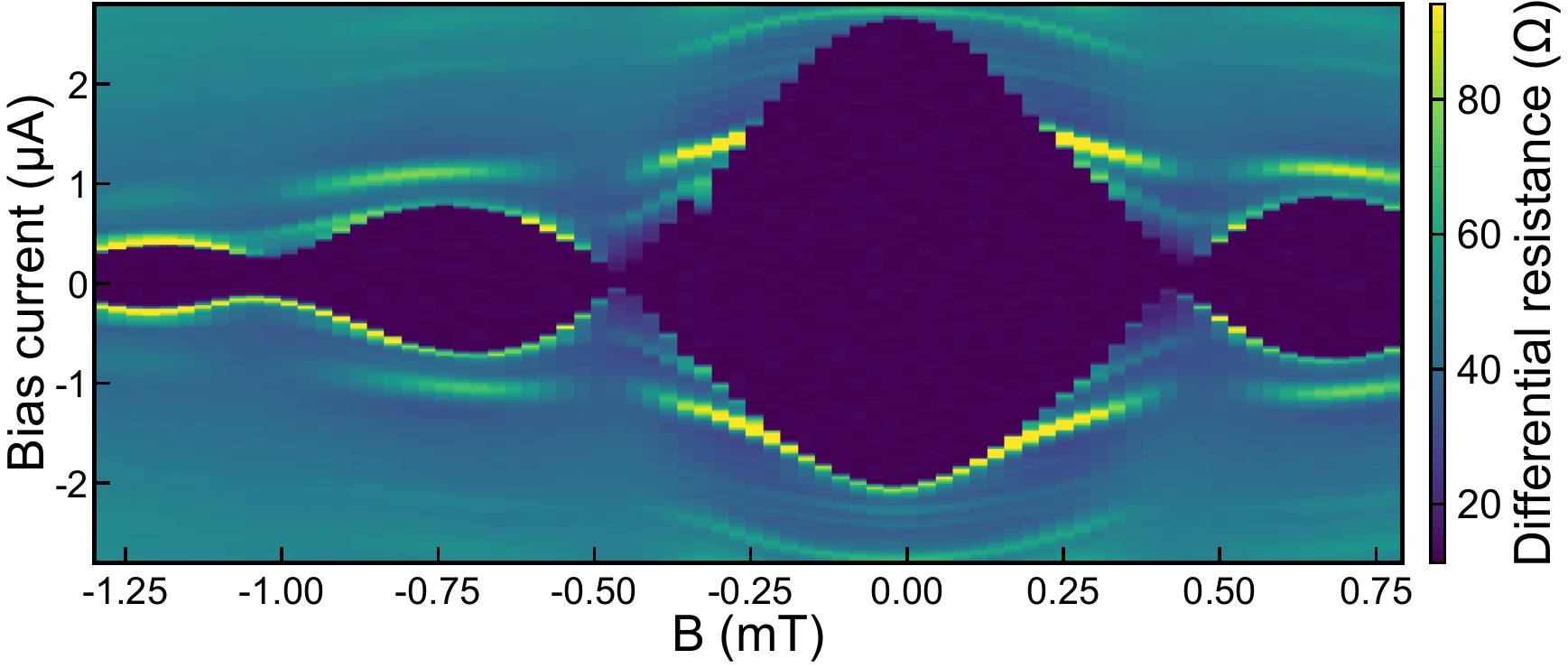}
\caption{Differential resistance of sample A as a function of the bias current and out-of-plane magnetic field. The critical current presents the typical Fraunhofer pattern of a JJ in the presence of an out-of-plane magnetic field.}
\label{figure4:Figure 4}
\end{figure}

By applying an external magnetic field to the junction, the critical current of the junction is found to follow a characteristic Fraunhofer pattern. As illustrated in Fig.~4 for sample A, the differential resistance oscillates as a function of the out-of-plane magnetic field. The pattern is periodic with a double-width central lobe, which suggests that the current is distributed uniformly in the junction and is not affected by distortion or wrinkles in the h-BN dielectric. We observe a slight increase of the period with the magnetic field which we attribute to the reduction of the field focusing effect caused by the Al contacts. Similar observations have been reported in Ref.~\cite{Suom} and can be attributed to the large width of our contacts.

In conclusion, we have demonstrated a versatile fabrication of thin h-BN flakes with Al-InAs JJ-FET with h-BN/graphite gate stack. These JJ-FET devices with h-BN dielectric exhibit equally pristine Josephson properties when compared to JJ-FET with Al$_2$O$_3$ dielectric. Through complimentary Hall bar studies we show that while Al$_2$O$_3$ deposition process increases the density of the 2DEG channel, the h-BN transfer is introducing less dopants. The small foot-print of the process and its flexibility in positioning the flakes together with minimal perturbation of the 2DEG density makes these new devices useful for applications in superconducting logic \cite{wen_josephson_2019} and quantum information technologies.

\textbf {Acknowledgment}  - NYU team acknowledges support from Army Research Office Agreements W911NF1810067 and W911NF1810115. Joseph Yuan acknowledges funding from the QUACRG (BAA W911NF-17-S-0002). Arkansas team acknowledges support from NSF under awards DMR-1610126 and DMR-1848281. K.W. and T.T. acknowledge support from the Elemental Strategy Initiative conducted by the MEXT, Japan, Grant Number JPMXP0112101001, JSPS KAKENHI Grant Numbers JP20H00354 and the CREST(JPMJCR15F3), JST.


\bibliography{References_Shabani_Growth}

\end{document}